\documentclass[11pt]{article}
\usepackage{moriond,epsfig}

\def\be{\begin{equation}}
\def\ee{\end{equation}}
\def\bea{\begin{eqnarray}}
\def\eea{\end{eqnarray}}

\newcommand{\zll}  {\mbox{${\mathrm Z}\rightarrow l^+ l^-$}}

\newcommand{\ztautau}  {\mbox{${\mathrm Z}\rightarrow\tau^+ \tau^-$}}

\newcommand{\wtaunu}  {\mbox{${\mathrm W}\rightarrow\tau \nu$}}

\newcommand{\zmumu}  {\mbox{${\mathrm Z}\rightarrow\mu^+ \mu^-$}}
\newcommand{\zee}  {\mbox{${\mathrm Z}\rightarrow e^+ e^-$}}

\newcommand{\wmunu}  {\mbox{${\mathrm W}\rightarrow\mu\nu$}}

\newcommand{\wenu}  {\mbox{${\mathrm W}\rightarrow{\mathrm e}\nu$}}

\newcommand{\met}  {\mbox{$\not{\! \! \! E_{\rm T}}$}}

\begin{document}
\vspace*{4cm}
\title{W and Z (plus Jets) Cross Sections at 1.96 TeV}

\author{ G. Hesketh\footnote{For the D\O\ and CDF collaborations} }

\address{Physics Department, Northeastern University, 111 Dana Research Center, Boston, MA, 02115}

\maketitle\abstracts{
We present an update on the status of the W and Z cross sections in proton-antiproton collisions at 1.96~TeV on behalf of the D\O\ and CDF collaborations.
Preliminary studies of the differential cross sections with different jet multiplicities and as a function of jet transverse energy are also presented.}

\section{W and Z Physics in Proton Anti-proton Collisions}
The electroweak program at the Tevatron is central to the aims of Run II, with the top and W mass and the W width measurements constraining the Standard Model higgs boson and beyond the Standard Model physics.
With the large integrated luminosities expected in Run II, the W and Z can also be used as probes of QCD.
Leptonic decay modes provide a clean experimental signal free from any final state hadronic interactions, and with 2~fb$^{-1}$ (expected  around 2007) samples of millions of Ws and hundreds of thousands of Zs will be available in each lepton channel.
Exclusive W and Z cross sections with different jet multiplicities are an excellent test of perturbative QCD, with the ratio of W or Z + $n$\ to W or Z + ($n+1$) jet events allowing a measurement of $\alpha_S$\cite{jet_ratio}.
Differential cross sections in terms of jet momentum and pseudo-rapidity allow tests of higher order QCD calculations\cite{cdf_wjets}. 
Production of W with jets is also one of the main backgrounds to top and Standard Model higgs physics in Run II, so a detailed understanding of these events is essential.
The Z boson makes an excellent QCD probe, as it can be precisely reconstructed and has little background.
Studies of Z transverse momentum will also feed into the understanding of hadronic recoil and can be used to calibrate the jet energy scale, replacing the photon plus jet events used in Run I\cite{d0_jes}. 
The jet energy scale is an important systematic in the W and top mass measurements, as well as QCD studies.

The forward backward asymmetries in W and Z events can be used to probe parton distribution functions and are sensitive physics beyond the Standard Model coupling to the W and Z bosons\cite{cdf_wasym,cdf_zasym}.
Extensions to the angular acceptance for leptons made possible by the Run II upgrades to the D\O\ and CDF detector improve the sensitivity of this measurement.

Overall, the electroweak program for Run II covers many exciting topics, and the first step on the way to these challenging measurements is the inclusive W and Z production cross section measurements.
These test our understanding of boson production (dominated by quark anti-quark annihilation), but also provide an excellent calibration source for detector performance.
From the W and Z cross section ratios, it is also possible to extract an indirect measurement of the W width\cite{d0_wwidth}, and preliminary results are presented here.

\section{Inclusive Cross Sections}
\subsection{Inclusive \zll\ Cross Sections}
Event selection for the Z cross section involves identifying two high energy leptons.
This signature also provides some redundancy, and in Z events identified by the presence of one lepton, it is possible to use the other lepton to measure all of the trigger and identification efficiencies needed for the Z and W analyses. 
Backgrounds to the Z are low, and come from two main sources.
The largest, \ztautau, is estimated with Monte Carlo and is typically less than 1\%.
Semi-leptonic quark decays are reduced using isolation requirements on the leptons, and the remaining contribution (less than 0.7\%) is estimated using the isolation and mass distributions in data.

\begin{figure}[ht]
\begin{center}
\psfig{figure=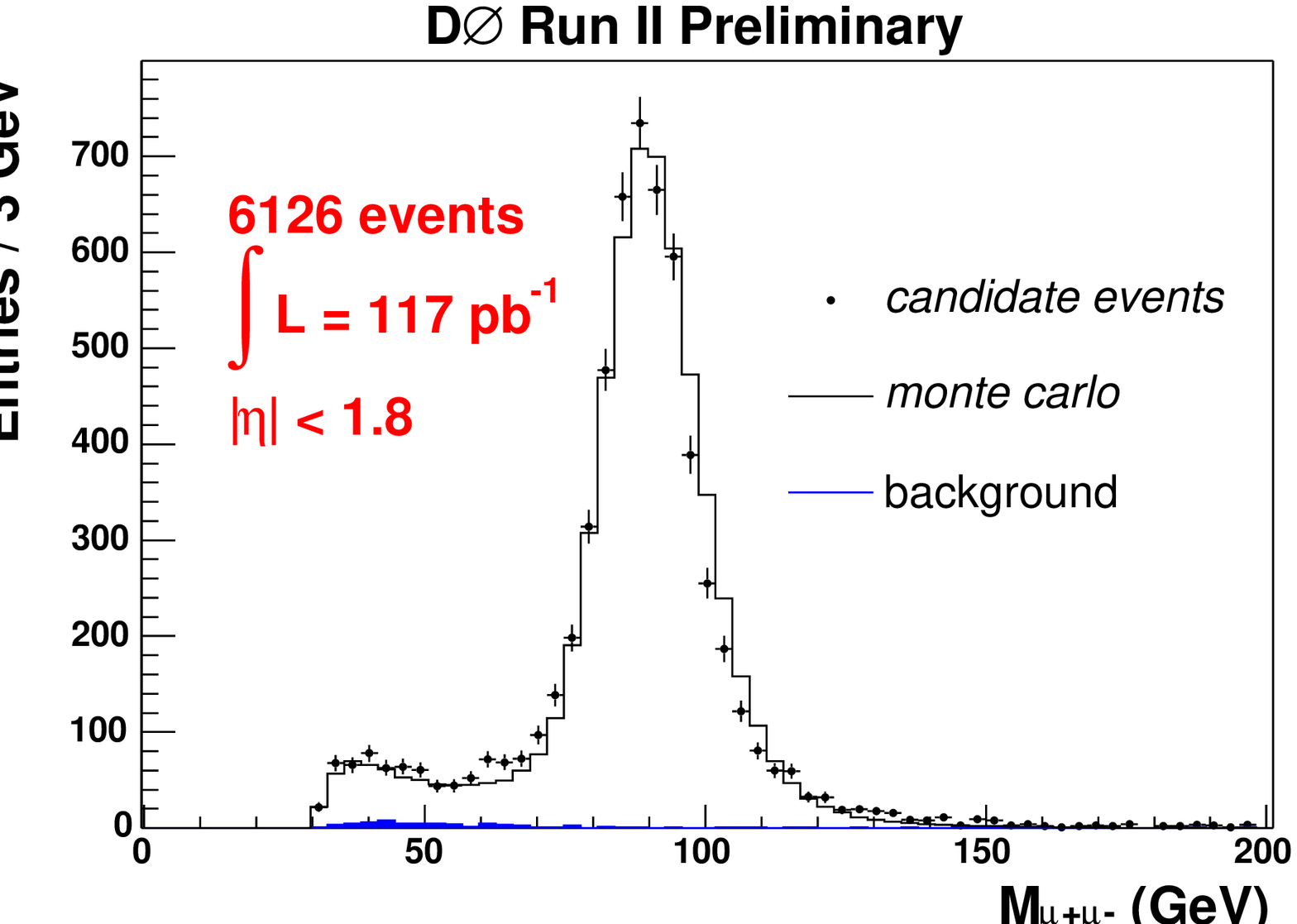,height=1.5in}
\psfig{figure=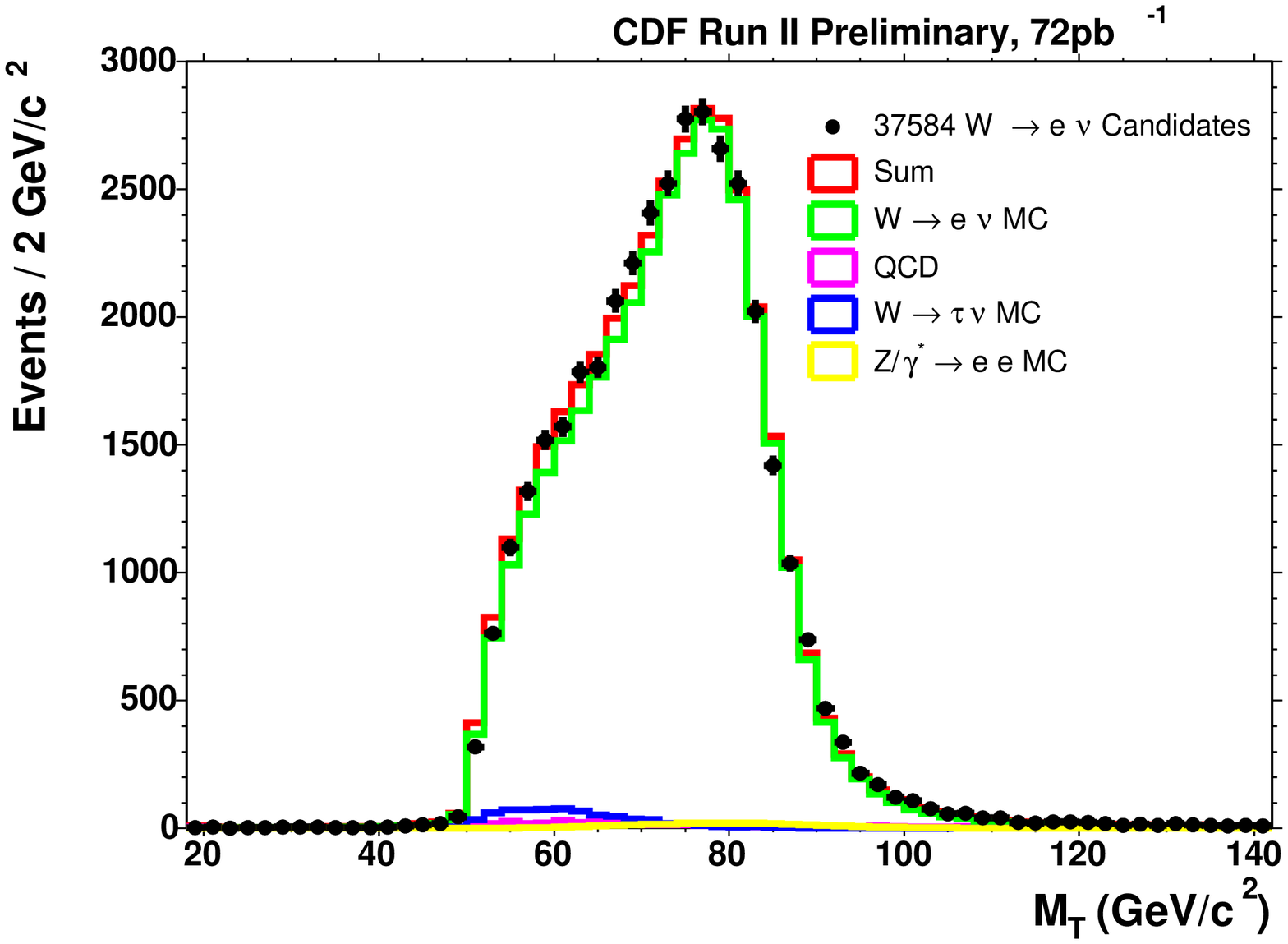,height=1.5in}
\end{center}
\caption{The di-muon mass (left) resulting from the Z cross section analysis at D\O, and the transverse mass (right) from the \wenu\ analysis at CDF.
\label{fig:zmass}}
\end{figure}

In order to extract the pure Z contribution to the measured Z/$\gamma^*$\ cross section, two methods are used.
At D\O, a broad mass range is considered (shown in figure \ref{fig:zmass}), and a correction is derived by comparing pure Z and Z/$\gamma^*$\ Monte Carlo.
At CDF, a narrower mass region is chosen (66-116~GeV) in which this correction is approximately unity.

The results obtained are:
\begin{itemize}
\item $\sigma(\mathrm{Z + X}).\mathrm{Br(\zmumu)} = 249 \pm 6(stat.) ^{+7}_{-6}(syst.) \pm 15(lum)$~pb, using 72~pb$^{-1}$\ (CDF),
\item $\sigma(\mathrm{Z + X}).\mathrm{Br(\zmumu)} = 262 \pm 5(stat.) \pm 9(syst.) \pm 26(lum)$~pb, using 117~pb$^{-1}$\ (D\O),
\item $\sigma(\mathrm{Z + X}).\mathrm{Br(\zee)} = 255 \pm 4(stat.) \pm 5(syst.) \pm 15(lum)$~pb, using 72~pb$^{-1}$\ (CDF),
\item $\sigma(\mathrm{Z + X}).\mathrm{Br(\zee)} = 275 \pm 9(stat.) \pm 9(syst.) \pm 29(lum)$~pb, using 42~pb$^{-1}$\ (D\O).
\end{itemize}

All results are in excellent agreement with the NNLO prediction of $252 \pm 9$~pb \cite{neerven}.
The uncertainty in the measured luminosity dominates all analyses, but other main systematics come from the uncertainties on the lepton identification efficiencies (less than 3\%), which will improve with more statistics, and the parton distribution functions (less than 1.5\%).

\subsection{Inclusive W Cross Sections}
In the case of \wmunu\ and \wenu, the neutrino is identified as missing transverse energy (\met) in an event, with the W event selection based on one high energy lepton and high \met. 
Compared to the Z, the W channels have higher statistics, but with only one reconstructed lepton, suffer higher backgrounds.
In both electron and muon modes, dominant backgrounds are semi-leptonic quark decays and \wtaunu.
As with the Z events, the semi-leptonic decays are reduced by isolation requirements on the reconstructed lepton, with the remaining contribution estimated in data to be around 1-2\%. 
The background due to \wtaunu\ is estimated using Monte Carlo to be around 2\%.
In the muon channel, the background due to \zmumu\ events (in which one muon lies outside the detector acceptance) is also significant, and is estimated in Monte Carlo to be around 5\%.
In the electron channel, with larger detector acceptance, this background is around 0.8\%.

The results obtained are:
\begin{itemize}
\item $\sigma(\mathrm{W + X}).\mathrm{Br(\wenu)} = 2782 \pm 14(stat.) ^{+61}_{-56}(syst.) \pm 167(lum)$~pb, using 72pb$^{-1}$\ (CDF),
\item $\sigma(\mathrm{W + X}).\mathrm{Br(\wenu)} = 2844 \pm 21(stat.) \pm 128(syst.) \pm 284(lum)$~pb, using 42pb$^{-1}$\ (D\O),
\item $\sigma(\mathrm{W + X}).\mathrm{Br(\wmunu)} = 2772 \pm 16(stat.) ^{+64}_{-60}(syst.) \pm 166(lum)$~pb, using 72pb$^{-1}$\ (CDF),
\item $\sigma(\mathrm{W + X}).\mathrm{Br(\wmunu)} =3226 \pm 128(stat.) \pm 100(syst.) \pm 322(lum)$~pb, using 18pb$^{-1}$\ (D\O),
\end{itemize}
to be compared with a NNLO prediction of $2690 \pm 100$~pb \cite{neerven}. 
Figure \ref{fig:zmass} shows the transverse mass distribution for the electron analysis at CDF, with the contributions from signal and background.
The dominant systematics in this analysis, after the luminosity uncertainty, come from the parton distribution function uncertainties (less than 1.5\%), lepton identification efficiencies and (for the electron channel at CDF) the material model used in the detector simulation (0.7\%).

\section{Cross Section Ratio and W Width}
In taking the ratio of the inclusive cross sections, ${\cal R}$, the largest uncertainty (from the luminosity determination) cancels, and there is partial cancellation of other systematic uncertainties, such as the lepton identification and parton distribution function uncertainty. 
CDF obtains preliminary results for this ratio in the electron channel of ${\cal R} = 10.86 \pm 0.18(stat) \pm 0.16(syst)$\ and in the muon channel of ${\cal R} = 11.10 \pm 0.27(stat) \pm 0.17(syst)$.
Combining these values gives ${\cal R} = 10.93 \pm 0.15(stat) \pm 0.14(syst)$.
In the electron channel, D\O\ obtains ${\cal R} = 10.35 \pm 0.35(stat) \pm 0.48(syst)$.

It is also possible to extract an indirect measurement of the W width, by expressing the ratio as:
\begin{equation}
  {\cal R} = \frac{\sigma_W}{\sigma_Z} \times \frac{\Gamma_Z}{\Gamma(\mathrm{Z} \rightarrow \mu\mu)} \times \frac{\Gamma(\mathrm{W} \rightarrow \mu\nu)}{\Gamma_W}
\end{equation}

Taking values for the ratio of absolute cross sections and W branching fractions from theory and the world average values for the Z width and branching fractions, the W width is the only unknown.
From the latest CDF results, a value of $\Gamma_W = 2071 \pm 40$~MeV is obtained, in good agreement with the Standard Model prediction of $2092 \pm 2.5$~MeV and a smaller uncertainty than the current world average of $2118 \pm 42$~MeV\cite{pdg}.
The Tevatron electro-weak working group will combine the CDF results with updated D\O\ results in the near future.

\section{W and Z Plus Jets, and Asymmetry}
Both CDF and D\O\ have made studies of jet multiplicities in W events.
To simulate W+jets production, the ALPGEN generator is used, achieving good agreement between \wenu\ data and Monte Carlo for the decrease in cross section with increasing jet multiplicities at both CDF and D\O. 
At jet multiplicities of three and higher, the background from top pair production dominates over the W plus jets signal.

\begin{figure}[ht]
\begin{center}
\psfig{figure=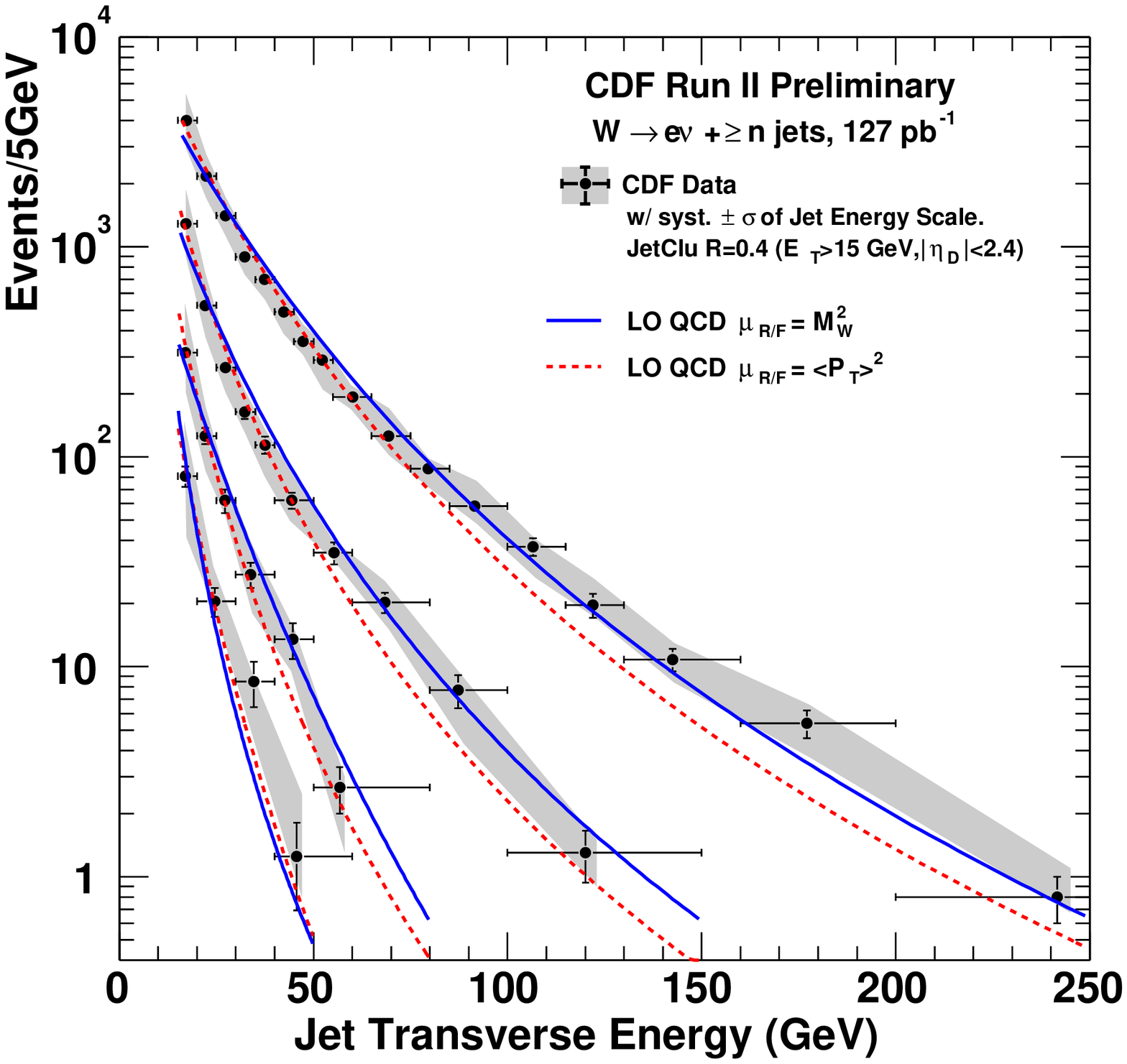,height=2in}
\psfig{figure=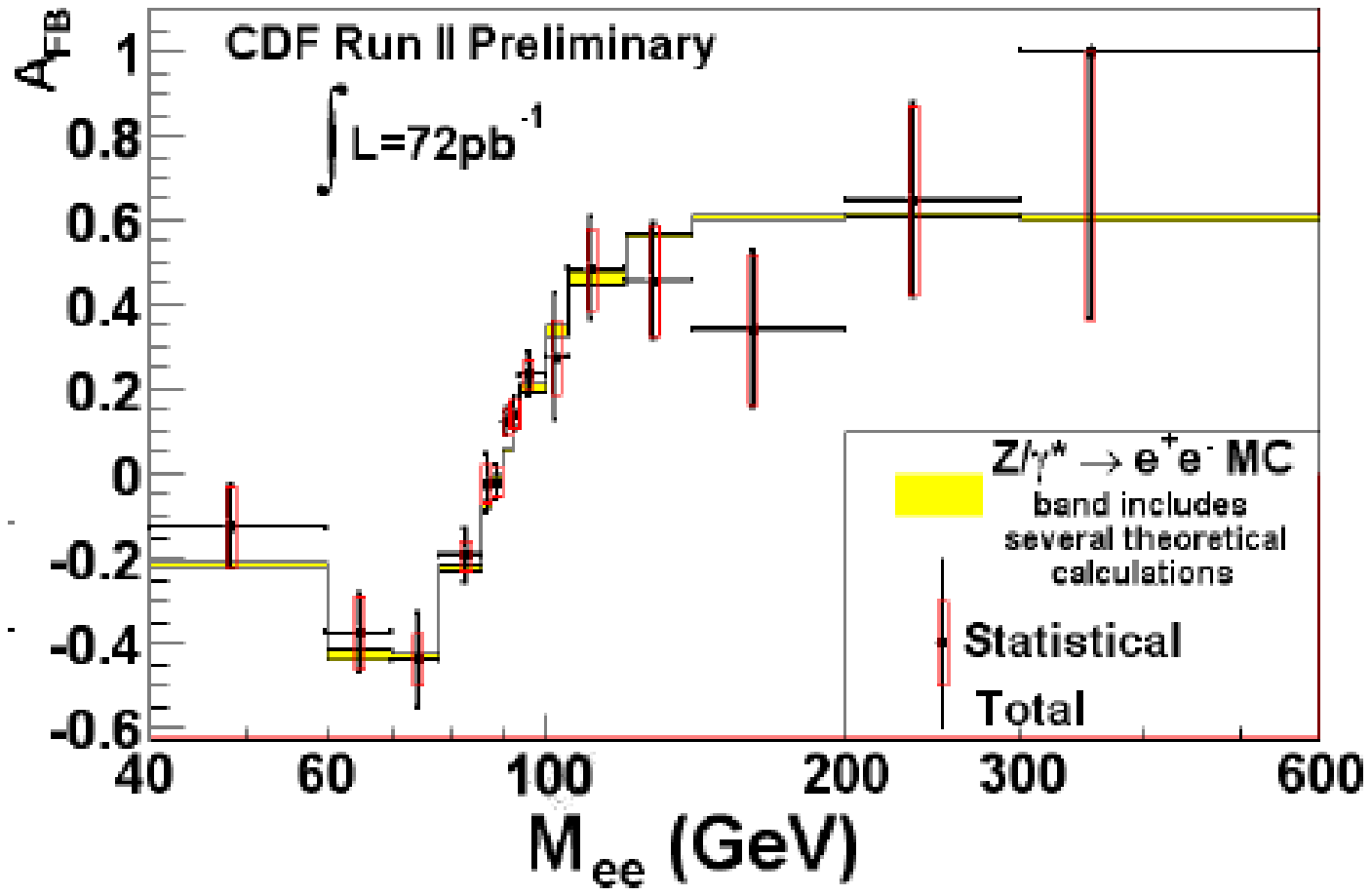,height=2in}
\end{center}
\caption{Jet transverse energy distributions (left) in data and Monte Carlo \wenu\ events, and the froward-backward asymmetry as a function of di-electron mass, both from CDF.
\label{cdf_jet}}
\end{figure}

CDF have made preliminary studies of the agreement between data and Monte Carlo at different jet energies, and with different jet multiplicities. 
Each line on figure \ref{cdf_jet} represents the transverse momentum of the $n^{th}$\ jet in W + $\geq n$\ jet events. 
Reasonable agreement can be seen over a wide range of jet momenta.
The uncertainties in data are dominated by the jet energy scale, shown as a grey band in figure \ref{cdf_jet}.
The di-jet mass in events with at least two jets is sensitive to the production mechanism and angular distribution of jets, and good agreement is found between data and Monte Carlo in a study by CDF.

The forward backward asymmetry in Z events is sensitive to couplings, and can be used to measure either the weak mixing anlge, or the ratio of up and down type quarks in the proton.
The asymmetry is also sensitive to new physics, through anomalous couplings.
CDF have carried out a preliminary study using central electrons, measuring the asymmetry as a function of di-electron mass.
This is shown in figure \ref{cdf_jet}, where 
there is good agreement between the data and Standard Model Monte Carlo.
Studies at D\O, which has a larger angular acceptance for leptons, are ongoing.
 
The studies of jet production in W events will also be possible in Z events, with the advantage of lower backgrounds and the ability to fully reconstruct the Z. 
This allows additional studies, such as the Z transverse momentum and rapidity, which can be used to tune simulation and parton distribution functions.
However, as the Z production cross section multiplied by branching fraction is an order of magnitude lower for the Z compared to the W, more statistics are needed and though the studies are ongoing, no results are presented here.

\end{document}